\documentclass[5p, number]{elsarticle}
\usepackage{flushend}
\usepackage{lipsum}
\usepackage{mathtools}
\usepackage{cuted}
\usepackage{float}
\usepackage{graphicx}
\graphicspath{ {./Images/} }
\usepackage{cite}
\usepackage[super]{nth}
\usepackage{amsmath,amssymb}
\usepackage{xcolor}
\usepackage{comment}
\makeatletter
\def\ps@pprintTitle{%
 \let\@oddhead\@empty
 \let\@evenhead\@empty
 \def\@oddfoot{}%
 \let\@evenfoot\@oddfoot}
\makeatother

\begin{document}

\begin{frontmatter}

\title{\Large {\sf Unified Mapping of Multi-Site Electrocatalytic Activity Using a Single  Descriptor}}

\author[MSE_CMU]{A.~Dana\corref{cor1}}
\ead{adana@andrew.cmu.edu}
\author[Rice]{D.~Terrones}
\author[MSE_CMU]{S.~Gelin\corref{cor1}}
\ead{sgelin@andrew.cmu.edu}
\author[MSE_CMU]{I.~Dabo\corref{cor1}}
\ead{idabo@andrew.cmu.edu}
\cortext[cor1]{Corresponding authors}

\address[MSE_CMU]{Department of Materials Science and Engineering, Carnegie Mellon University, Pittsburgh, PA 15213, USA.}
\address[Rice]{Department of Physics, Rice University, Houston, TX 77005, USA.}

\begin{abstract} We present a precise and general method to map the activity of electrocatalysts across multiple  sites. Starting from a mean-field statistical mechanics model, we introduce an effective adsorption free energy descriptor that explicitly incorporates lateral adsorbate–adsorbate interactions, enabling the construction of coverage-consistent volcano relationships. Extending this approach, we show that adsorption energetics and interaction strength define a two-dimensional activity landscape that gives rise to a `volcano ridge' that captures the coupled influence of binding and interactions on catalytic performance. For multi-site systems, we demonstrate that the inherently nonlinear coupling between distinct adsorption environments leads to multi-peaked activity trends that cannot be represented by conventional single-site descriptors. To address this, we introduce a reduced descriptor mapping that projects the multidimensional activity landscape onto a single effective coordinate while preserving the underlying physics of site heterogeneity and lateral interactions. The resulting framework generalizes Sabatier-type analysis to complex alloy catalysts and provides a physically interpretable route for screening electrocatalytic materials of arbitrary compositional complexity.
\end{abstract}

\end{frontmatter}

\section{Introduction}

\begin{figure*}[ht]
    \centering
    \includegraphics[width=\textwidth]{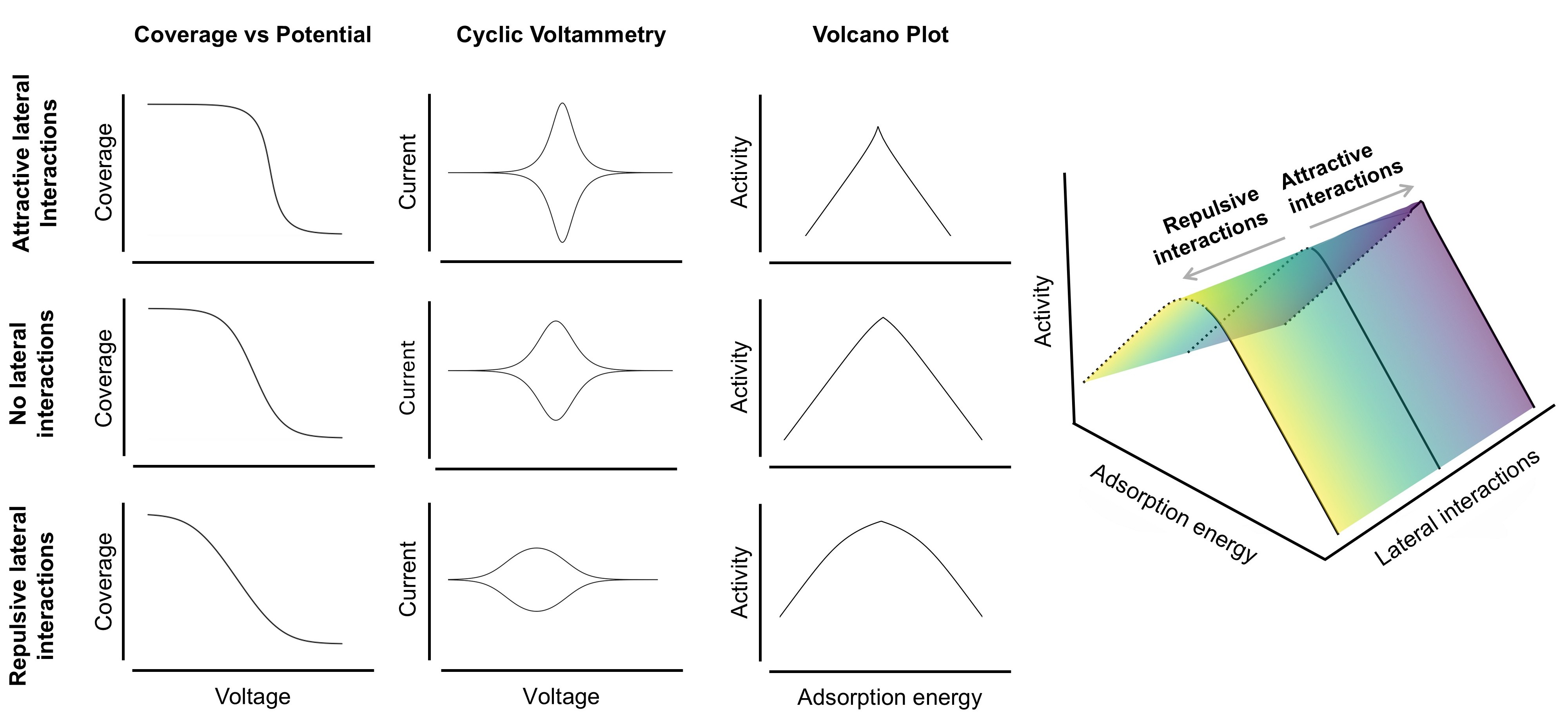}
    \caption{Effect of lateral adsorbate-adsorbate interactions on volcano plot and electrochemical observables. Repulsive interactions flatten the volcano peak, whereas attractive interactions narrow the peak, resulting in a three-dimensional volcano depiction with lateral interactions as the third dimension. For the cyclic voltammetry (CV) response, repulsive interactions shift the CV peak to lower potentials and reduce its intensity. Conversely, attractive interactions result in the opposite pattern. The coverage-potential curve becomes steeper for attractive interactions, while repulsive interactions cause a more gradual shift.}
    \label{fig1}
\end{figure*}

Volcano curves are widely used to assess the performance of electrocatalysts, offering a simple depiction of activity trends rooted in Sabatier's principle \citep{Sabatier1913, Sabatier1911HydrognationsED}. While this framework has been successful for pure metals and structurally simple surfaces, its applicability becomes increasingly limited for more complex catalytic substrates. Two fundamental challenges arise when extending conventional volcano models to realistic systems. First, adsorption energies are inherently coverage-dependent due to lateral adsorbate–adsorbate interactions, introducing ambiguity in the choice of descriptor values and leading to shifts in predicted activity trends. Second, alloy surfaces exhibit multiple, chemically distinct adsorption sites, resulting in a distribution of adsorption energetics rather than a single well-defined binding energy. These effects are further coupled through the dependence of surface coverage on electrode potential, which dynamically modifies both adsorption energetics and interaction strengths. As a result, the central assumption of a single, coverage-independent descriptor for conventional volcano plots breaks down.

To put this problem into perspective, Fig.~\ref{fig1} illustrates how lateral adsorbate-adsorbate interactions modify adsorption thermodynamics and, in turn, catalytic activity trends. These interactions affect the adsorption isotherm where attractive interactions lead to a steeper adsorption isotherm, while repulsive interactions produce a more gradual variation of the adsorption isotherm. The altered adsorption isotherm directly influences the voltammetric response. Attractive interactions shift voltammetric peaks toward higher potentials and increase their intensity, while repulsive interactions shift the peaks toward lower potentials and reduce their intensity. When incorporated into volcano analysis, these coverage effects translate into systematic changes in activity trends, with repulsive interactions broadening and flattening the volcano peak and attractive interactions narrowing the peak. This approach extends the conventional volcano concept to a generalized representation, in which a third dimension captures the strength of lateral interactions, highlighting their fundamental role in electrocatalysis.

Several attempts have been made to construct generalized volcano curves \citep{chen2024restructuring, liang2024unravelling, ke2022three,grabow2010understanding, wang2025revisiting, qi2012adsorbate,yang2022reconciling}. Yang \textit{et al.}~\citep{yang2022reconciling} utilized the Butler--Volmer relationship to depict volcano trends using different transfer coefficients. Liang \textit{et al.}~\citep{liang2024unravelling} have examined the effect of lateral interactions of $\mathrm{IrO}_x$ on $\mathrm{O_2}$ evolution. Qi \textit{et al.}~\citep{qi2012adsorbate} have used a mean-field microkinetic model to simulate the electrochemical oxygen reduction process on surfaces of Pt(111) and Pt(100) and report flattening of volcano peak due to lateral repulsions. Grabow \textit{et al.}~\citep{grabow2010understanding} investigated the interactions between adsorbates for CO oxidation over transition metals. There have also been studies that depict synergy-dependent volcano~\citep{chen2024restructuring}, three-dimensional volcano plot under an external electric field~\citep{ke2022three}, and the use of the Lambert W function to construct a three-dimensional volcano~\citep{wang2025revisiting}. These efforts have provided important insights into the influence of adsorbate interactions and reaction kinetics on the activity of single-site electrocatalysts. 

In this work, we develop a unified analytical framework that explicitly incorporates coverage-dependent adsorption energetics and lateral interactions within a single formalism applicable to multi-site catalysts. Building on a mean-field statistical mechanics description, we derive effective descriptors that capture the interplay between binding energetics and interaction effects, enabling the construction of generalized volcano relationships that remain valid across a wide range of surface coverages. HER provides an ideal test case for this analysis due to the extensive experimental and computational efforts devoted to identifying active catalysts. A broad range of alloying elements, including Co \citep{lin2020structurally, zhang2021engineering, yu2023high, zhang2019bimetallic}, Fe \citep{qiao2021pt3fe, zhong2017double}, Au \citep{fu2023continuous}, Ge \citep{mondal2022morphology}, Ni \citep{zhang2021wox}, Ru \citep{pang2022laser}, Sn \citep{chen2022ptco}, and Ir \citep{ni2023construction}, have been studied both experimentally and computationally for HER. In addition, volcano trends for HER on pure elements have been widely reported. Trasatti \textit{et al.}~\citep{trasatti1972work} constructed one of the earliest experimental volcano plots, while N{\o}rskov \textit{et al.}~\citep{norskov2005trends} provided an important theoretical description of activity trends.

We have validated our model by comparing the predicted coverage curves and HER activities with experimental results and then benchmarked the activity of different platinum-based alloys by introducing a three-dimensional volcano plot (volcano ridge) that accounts for both adsorption energies and lateral interactions. We conclude by proposing a reduced descriptor mapping that projects the multidimensional activity landscape onto a single effective coordinate, resulting in a multi-peaked volcano relationship that is applicable to catalytic alloys of any complexity while maintaining the conceptual simplicity of Sabatier's principle.

\section{Results}
\subsection{Modeling Pure Metals}
\subsubsection{Single-Site Mean-Field Adsorption Model}
 To compute the surface hydrogen coverage, we adopt a grand canonical ensemble framework, in which the adsorbed hydrogen atoms are in thermodynamic equilibrium with a reservoir of protons and electrons. Within this ensemble, the appropriate thermodynamic potential is the grand canonical potential $\Omega$, which accounts for energy and entropy contributions as well as the exchange of particles with external reservoirs and is defined as 
\begin{equation} 
\Omega = E - T S - \sum_i \mu_i N_i,
\label{eq:omega} \end{equation}
where $E$ denotes the energy of the adsorbed species, $T$ is the absolute temperature, $S$ is the entropy of the system, $N_i$ is the number of particles of species $i$, and $\mu_i$ is the corresponding chemical potential. The relevant particles exchanged with the reservoir are protons and electrons, whose chemical potentials are explicitly incorporated into the grand canonical formulation. The energy of the adsorbed species, $E$, is computed using density-functional theory (DFT) from the expression
\begin{equation} 
 E = E_{\rm slab}(N) -  E_{\rm slab}(0),
\label{eq:E_dft} 
\end{equation}
where $E_{\rm slab}(N)$ is the total energy of the slab with $N$ adsorbed hydrogen atoms and $E_{\rm slab}(0)$ is the energy of the slab without adsorbates. To describe the relationship between hydrogen adsorption energy and surface coverage, the lattice-based mean-field model is employed. In this approach, the adsorption sites on the metal surface are treated as a discrete lattice, where each site can be either occupied by a hydrogen atom or remain vacant. The total adsorption energy, $\Delta E$, is modeled using an effective Hamiltonian that includes both on-site adsorption energies and pairwise interactions between neighboring adsorbates. Specifically, the energy $\Delta E \equiv E - \tfrac{1}{2} N\mu^\circ_{\rm H_2}$, with $\mu^\circ_{\rm H_2}$ being the chemical potential of ${\rm H_2}$ in its standard state (see Methods), is approximated as 
\begin{equation} 
\Delta E = IN+\frac12 J N z \theta,
\label{eq:E_model_extended} 
\end{equation}
where $I$ represents the hydrogen binding energy of an isolated site, $J$ denotes the interaction energy between neighboring hydrogen atoms, $N$ is the number of hydrogen atoms adsorbed on the surface, $z$ is the number of nearest-neighbor sites available for interaction (determined by the lattice geometry), and $\theta$ is the hydrogen coverage defined as number of adsorbed hydrogen over total number of available adsorption sites ($\theta={N}/{N_{\rm s}}$). 

This formulation captures both the individual adsorption energy per hydrogen atom and the configurational effects arising from lateral interactions between neighboring adsorbates. The model provides a foundation for evaluating the thermodynamics of adsorption when combined with mean-field approximations.
Eq.~\ref{eq:E_model_extended} establishes a relationship between the adsorption energy and the surface coverage of hydrogen. The parameters $I$ (the isolated hydrogen binding energy) and $J$ (the hydrogen–hydrogen interaction energy) are obtained by fitting calculated adsorption energies to the model in Eq.~\ref{eq:E_model_extended} across a range of coverages and configurations. 

\begin{figure}[ht]
    \centering
    \includegraphics[width=\columnwidth]{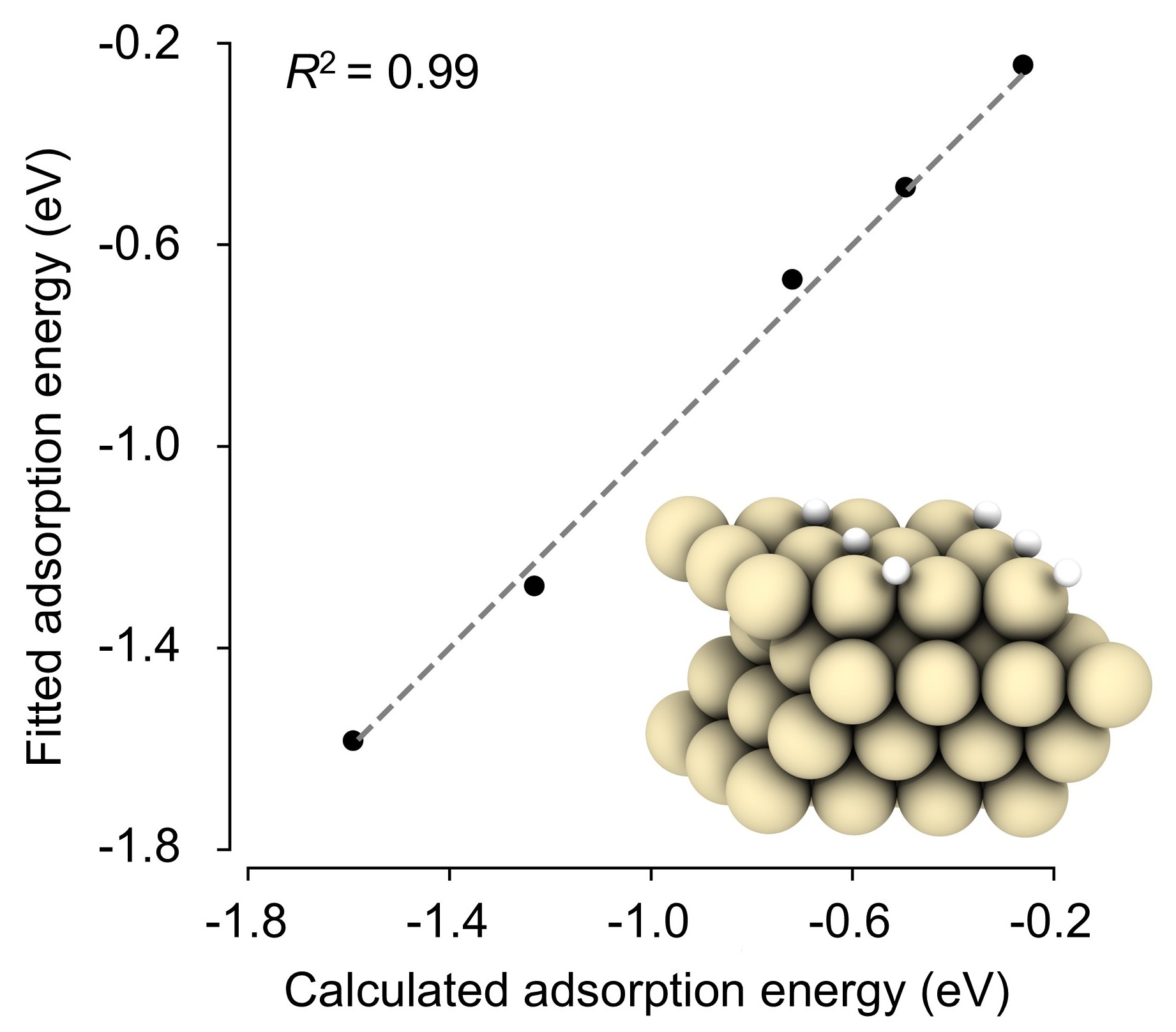}
    \caption{Parity plot comparing adsorption energies obtained from the coverage-dependent model with corresponding predicted values for Pt(111), including a representative monolayer with 6/12 coverage.}
    \label{fig2}
\end{figure}

A four-layer slab with twelve Pt atoms in each layer was constructed and hydrogen adsorption energies were computed at six distinct surface coverages ranging from 1/12 monolayer to full monolayer coverage ({Fig.~S1}), as shown in Fig.~\ref{fig2}. The hydrogens were placed on the fcc site, which was identified as the most energetically favorable adsorption site from electronic-structure calculations ({Fig.~S2a}). Adsorption energies were used to fit the parameters in Eq.~\ref{eq:E_model_extended}, and the predicted energies from the fitted model show excellent agreement with their calculated counterparts, as illustrated in Fig.~\ref{fig2}. The model achieves a high coefficient of determination, $R^2=0.99$, indicating its accuracy in reproducing the energetics of hydrogen adsorption on Pt.

\subsubsection{Coverage Isotherm}
To calculate the coverage response, the complete grand canonical ensemble framework must be considered, meaning the entropic contribution to the free energy must also be accounted for. Using a regular mixing model, the configurational entropy associated with adsorbate distributions on the surface is given by 
\begin{equation} 
S_{\rm conf} = - Nk_{\rm B}[\theta\ln{\theta}+(1-\theta)\ln{(1-\theta)}],
\label{eq:S} 
\end{equation}
where $k_{\rm B}$ is the Boltzmann constant. By combining Eq.~\ref{eq:omega} with Eq.~\ref{eq:E_model_extended} and Eq.~\ref{eq:S}, the grand canonical potential can be expressed as
\begin{multline}
\Omega = IN+\frac12 J N z \theta + Nk_{\rm B}T[\theta\ln{\theta}+(1-\theta)\ln{(1-\theta)}]\\ + Nk_{\rm B}T \ln(10)\,\mathrm{pH} + N e (U-U_{\rm SHE}).
\label{eq:omega_extended} 
\end{multline}

This expression for the grand canonical potential, $\Omega$, accounts for the energetic and entropic contributions associated with hydrogen adsorption on the surface, as well as the exchange of particles with proton and electron reservoirs. The first two terms represent the adsorption energy based on the lattice model, incorporating both site-specific binding and hydrogen--hydrogen interactions. The third term captures the configurational entropy of hydrogen adsorbates, treated within a mean-field approximation. The final two terms reflect the energetic cost of exchanging hydrogen atoms with the surrounding electrochemical environment through the chemical potentials of protons $\mu_{\mathrm{{\rm H}^+}}$ and that of the electrons $\mu({e^-})$ at the applied potential $U$ ($U_{\rm SHE}=\rm 4.44\,V$;     cf. Methods section for details).

To determine the equilibrium hydrogen coverage, the grand canonical potential $\Omega$ is minimized with respect to the number of adsorbed hydrogen atoms, $N$, at fixed temperature and chemical potentials. This condition is expressed as 
\begin{equation}
\frac{\partial\Omega}{\partial N} =0.
\end{equation}
Taking the derivative of Eq.~\ref{eq:omega_extended} with respect to $N$, at $\mathrm{pH}=0$, yields 
\begin{equation}
\frac{\partial\Omega}{\partial N} =I+J z \theta + k_{\rm B}T\ln{\frac{\theta}{1-\theta}}+ N e (U-U_{\rm SHE}).
\label{eq:omega/N}
\end{equation}
By setting the derivative to be equal to zero and rearranging terms, one obtains
\begin{equation} 
I+J z \theta + k_{\rm B}T\ln{\frac{\theta}{1-\theta}} =-e(U-U_{\rm SHE}).
\label{eq:final_Pt} 
\end{equation}
Equation \ref{eq:final_Pt} provides a direct relationship between the hydrogen coverage $\theta$ and the electrode potential $U$. By solving this equation, the equilibrium hydrogen coverage as a function of the applied electrode potential is determined. This coverage–potential relationship serves as a basis for computing key electrochemical indicators, such as catalytic activity. 

\begin{figure}[ht]
    \centering
    \includegraphics[width=\columnwidth]{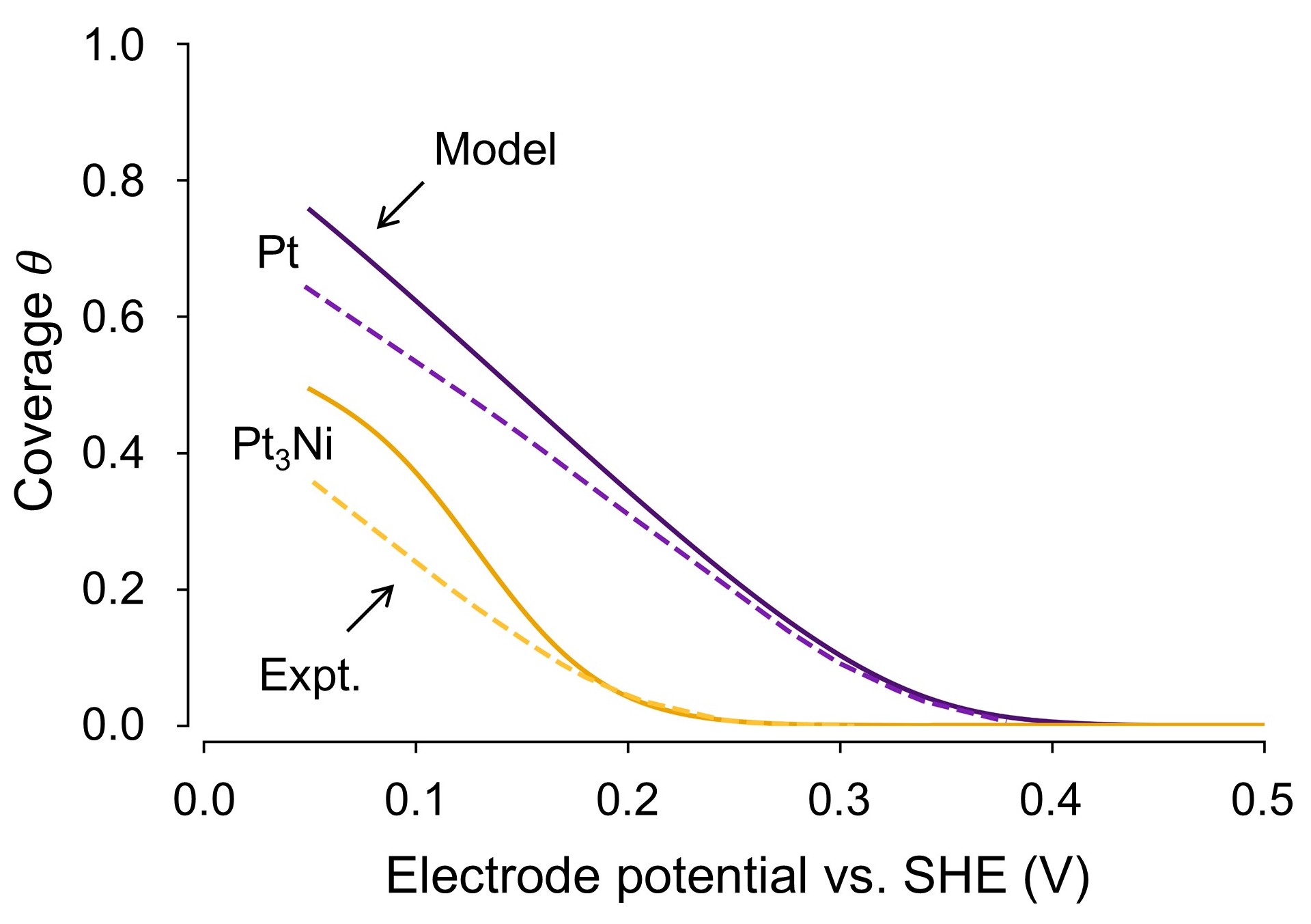}
    \caption{Hydrogen coverage as a function of electrode potential on Pt(111) and $\mathrm{Pt_3Ni(111)}$ surfaces (dashed lines represent experimental data and solid lines are model predictions). The consistency between model predictions and experimental coverage-potential measurements~\citep{stamenkovic2007improved} demonstrates that the model reliably accounts for adsorption energetics and lateral interactions on these surfaces.
    }
    \label{fig3}
\end{figure}

The accuracy of the model is evaluated against experimental coverage-potential data reported by Stamenkovic \textit{et al.} \citep{stamenkovic2007improved}. Hydrogen coverage on the Pt(111) surface as a function of applied potential is computed using Eq.\ref{eq:final_Pt}. {All fitted parameters used in the model are summarized in Table S1 in the Supplementary Information}. The resulting coverage curve is presented in Fig.~\ref{fig3} and compared with the experimental data from Stamenkovic\textit{ et al.} \citep{stamenkovic2007improved}. Our model shows close agreement with the experimental measurements, demonstrating its predictive accuracy.

\subsubsection{Volcano Plots}
Using the model presented above, we can be extract the activity of the catalyst in the form of the exchange current by using the method proposed by Yang \textit{et al.} \citep{yang2022reconciling}. In this model, hydrogen evolution proceeds via the following steps:
\begin{equation}
*+{\rm H}^+_{\rm (aq)} + e^- \leftrightarrows {\rm H}^*, 
\label{reaction_H*_1}
\end{equation}
\begin{equation}
{\rm H}^* \leftrightarrows \frac12{\rm H}_{2\rm (g)}.
\label{reaction_H2_1}
\end{equation}
If hydrogen adsorption is energetically favorable, Eq.~\ref{reaction_H2_1} becomes the rate-limiting step for HER. Conversely, when hydrogen adsorption on the metal surface is energetically unfavorable, Eq.~\ref{reaction_H*_1} is the rate-limiting step. The general expression of the electrochemical current at equilibrium for a single-step, single-electron transfer reaction is~\citep{yang2022reconciling, bard2022electrochemical}:
\begin{equation}
i_0 = e k_0 \Gamma_{+}^{1-\alpha} \Gamma_{-}^{\alpha},
\end{equation}
where $i_0$ denotes the exchange current density, $e$ is the electron charge, $k_0$ is the standard rate constant, and $\alpha$ is the charge transfer coefficient. $\Gamma_{+}$ and $\Gamma_{-}$ represent the surface concentrations of reactive sites for the forward and backward reactions, respectively.

For a water-metal interface operating outside diffusion limitations of H$^+$ and H$_2$, HER is governed primarily by the density of unoccupied adsorption sites $\Gamma_{*}$ and the coverage of adsorbed hydrogen $\Gamma_{\rm H^*}$, which sum up to total areal density of adsorption sites: 
\begin{equation}
\Gamma = \Gamma_{*} + \Gamma_{\rm H^*}
\end{equation}
where $\Gamma_{\rm H^*}=\Gamma \theta$ and $\Gamma_{\rm H^+}=\Gamma (1-\theta)$ for the HER steps in Eq.~\ref{reaction_H*_1} and Eq.~\ref{reaction_H2_1}. At equilibrium, the rates involved in the forward and backward steps must be the same, and the rate constant for getting out of the chemisorption state is considered independent of the metal.  When hydrogen adsorption is energetically favorable, meaning Eq.~\ref{reaction_H2_1} is the rate-limiting step, the exchange current density takes the following form:
\begin{equation}
i_0=-ek_0\Gamma(1-\theta)^\alpha\theta^{1-\alpha}.
\label{eq:i0}
\end{equation}
Conversely, when hydrogen adsorption energy is energetically unfavorable, meaning Eq.~\ref{reaction_H2_1} is the rate-limiting step, the exchange current density is given by the following equation:
\begin{equation}
i_0=-ek_0\Gamma(1-\theta)^{1-\alpha}\theta^{\alpha}
\label{eq:i0_1}
\end{equation}
where $k_0 = 100\ \mathrm{s^{-1}\,site^{-1}}$ is a fitted kinetic prefactor chosen to yield a realistic estimate of the reaction rate. The total areal density of adsorption sites $\Gamma$ is determined for each element based on the simulated structures, while the surface coverage $\theta$ is obtained from Eq.~\ref{eq:final_Pt} and Eq.~\ref{eq:final_PtM} for platinum and platinum-alloy catalysts, respectively. The transfer coefficient $\alpha$ is selected based on experimental values reported for pure elements by Yang \textit{et al.}~\citep{yang2022reconciling}. For the platinum alloys investigated in this study, a transfer coefficient of $\alpha = 0.5$ is assumed, in accordance with the value reported for platinum.

The criterion used to distinguish between exothermic and endothermic hydrogen adsorption is the hydrogen adsorption free energy at zero coverage, $\Delta G_{{\rm H}^*}(\theta=0)$, which determines whether Eq.~\ref{eq:i0} or Eq.~\ref{eq:i0_1} should be applied to calculate $j_0$, respectively:
\begin{equation}
\Delta G_{{\rm H}^*}=I+\Delta E_{\rm ZPE}-T\Delta S_{\rm H}.
\label{eq:volcano_G}
\end{equation}
Here, $\Delta E_{\rm ZPE}$ is the difference in zero-point energy between the adsorbed and the gas phase and $T\Delta S_{\rm H}$ is the entropy of $H_2$ in the gas phase at standard conditions. These values are taken from N{\o}rskov \textit{et al.} \citep{norskov2005trends} and are equal to 0.04 eV and 0.20 eV, respectively.

\begin{figure}[ht]
    \centering
    \includegraphics[width=\columnwidth]{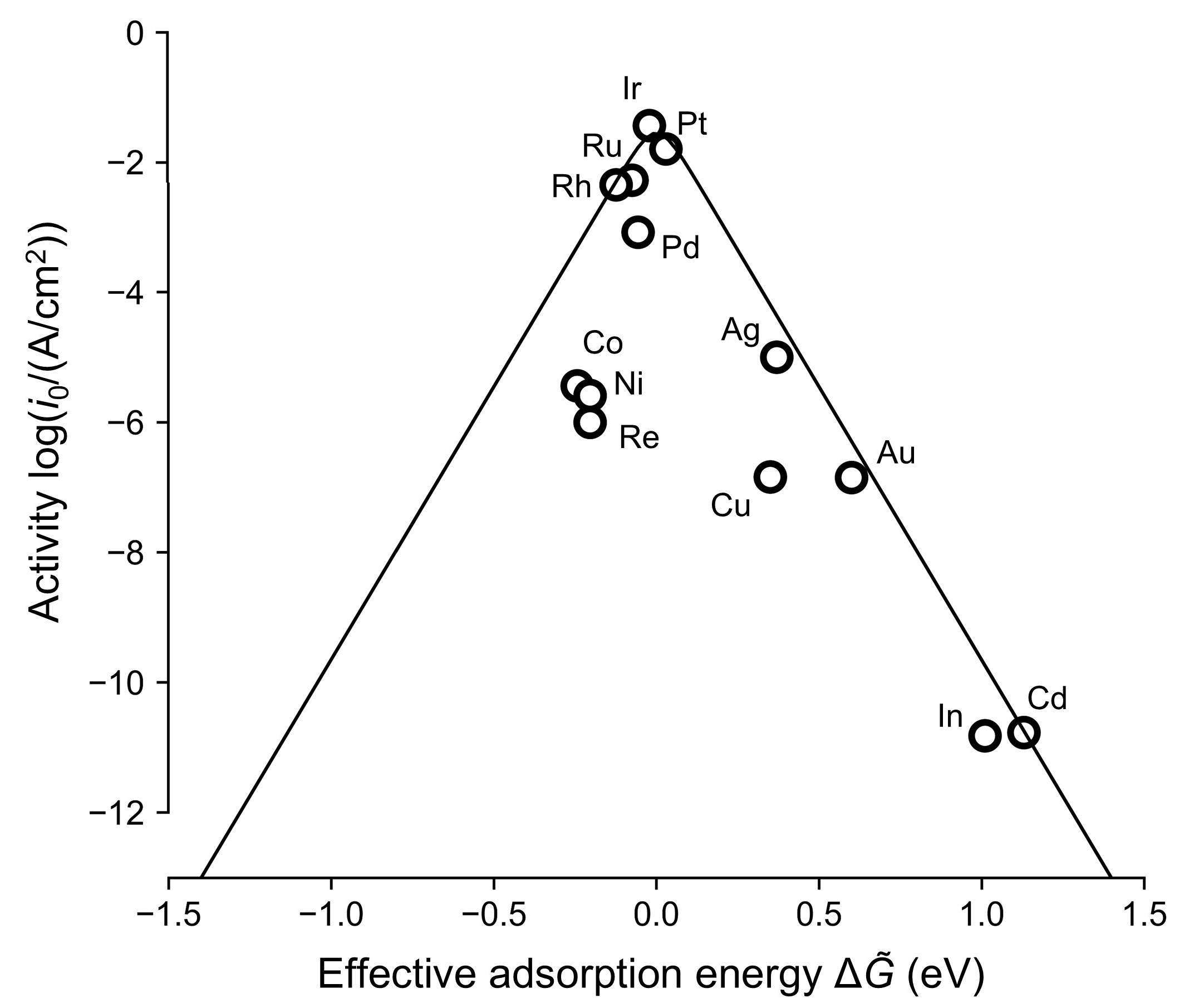}
    \caption{Simplified catalytic volcano plot constructed using the effective hydrogen adsorption free energy descriptor, ${\Delta \tilde G}_{\mathrm{H}^\ast}$, which incorporates both coverage dependence and lateral adsorbate-adsorbate interactions. This descriptor enables the unification of activity trends across different surface coverages into a single curve. The calculated volcano illustrates the relationship between the effective adsorption free energy and catalytic activity, demonstrating how the inclusion of interaction effects modifies the position and shape of the activity maximum. The volcano is obtained using a transfer coefficient of $\alpha = 0.5$.}
    \label{fig4}
\end{figure}

 A significant challenge in constructing volcano plots for electrocatalytic activity is the intrinsic coverage dependence of adsorption energies, which produces distinct activity trends at varying surface coverages. By changing the coverage of adsorbates, the calculated adsorption energy changes which will cause a shift in the trends.  Consequently, a single catalyst can display multiple volcano curves as a function of hydrogen coverage.  
 To address the explicit coverage dependence of the activity descriptor, an effective hydrogen adsorption free energy is proposed that accounts for both coverage effects and lateral interactions. Based on Eq.~\ref{eq:final_Pt}, this effective descriptor is defined as:
\begin{equation}
\Delta \tilde{G}_{\mathrm{H}^*}
= I + z \theta J + 0.24 \ \text{eV}.
\label{eq:activity2}
\end{equation}
Here, ${\Delta \tilde G}_{\mathrm{H}^\ast}$ serves as an effective descriptor for constructing simplified volcano plots, allowing activity patterns from different coverages and lateral interactions to be combined into a single curve. To demonstrate the effectiveness of ${\Delta \tilde G}_{\mathrm{H}^\ast}$ as an effective descriptor, we used adsorption energy data reported by Yang \textit{et al.}~\citep{yang2022reconciling} to extract the corresponding $I$ and $J$ parameters for each element. These parameters are then used to compute surface coverages, from which ${\Delta \tilde G}_{\mathrm{H}^\ast}$ is calculated. The resulting descriptor is then plotted against experimentally measured activities in Fig.~\ref{fig4}. This comparison demonstrates that the proposed descriptor successfully reproduces the characteristic volcano-like trend, highlighting its ability to connect theoretical energetics with experimentally measured activities.

\subsection{Modeling Alloys}
\subsubsection{Multi-Site Mean-Field Adsorption Model}
To study bimetallic surfaces, a four-layer slab was constructed to represent the PtNi alloy surface, where only the second atomic layer contains 50\% Ni, following Ref.~\citep{stamenkovic2007trends},. This structural model reflects the experimental observations reported by Stamenkovic \textit{et al.} \citep{stamenkovic2007trends}, in which the topmost surface layer is composed entirely of Pt, the second layer contains 50\% Ni, and the third layer is enriched with 87\% Pt, with subsequent layers following the bulk composition. For computational efficiency and minimal loss of accuracy, the alloy representation was simplified: the solute atoms were restricted to the second layer only, as contributions from Ni atoms in the third layer are expected to be negligible due to their depth, and the impact of 13\% Ni in the third layer is marginal with only 12 atoms in each layer. All symmetrically distinct configurations with 50\% Pt and 50\% Ni in the second layer were examined, and the structure with the lowest total energy was selected as the representative slab ({Figs.~S3 and S4}). 

\begin{figure}[ht]
    \centering
    \includegraphics[width=\columnwidth]{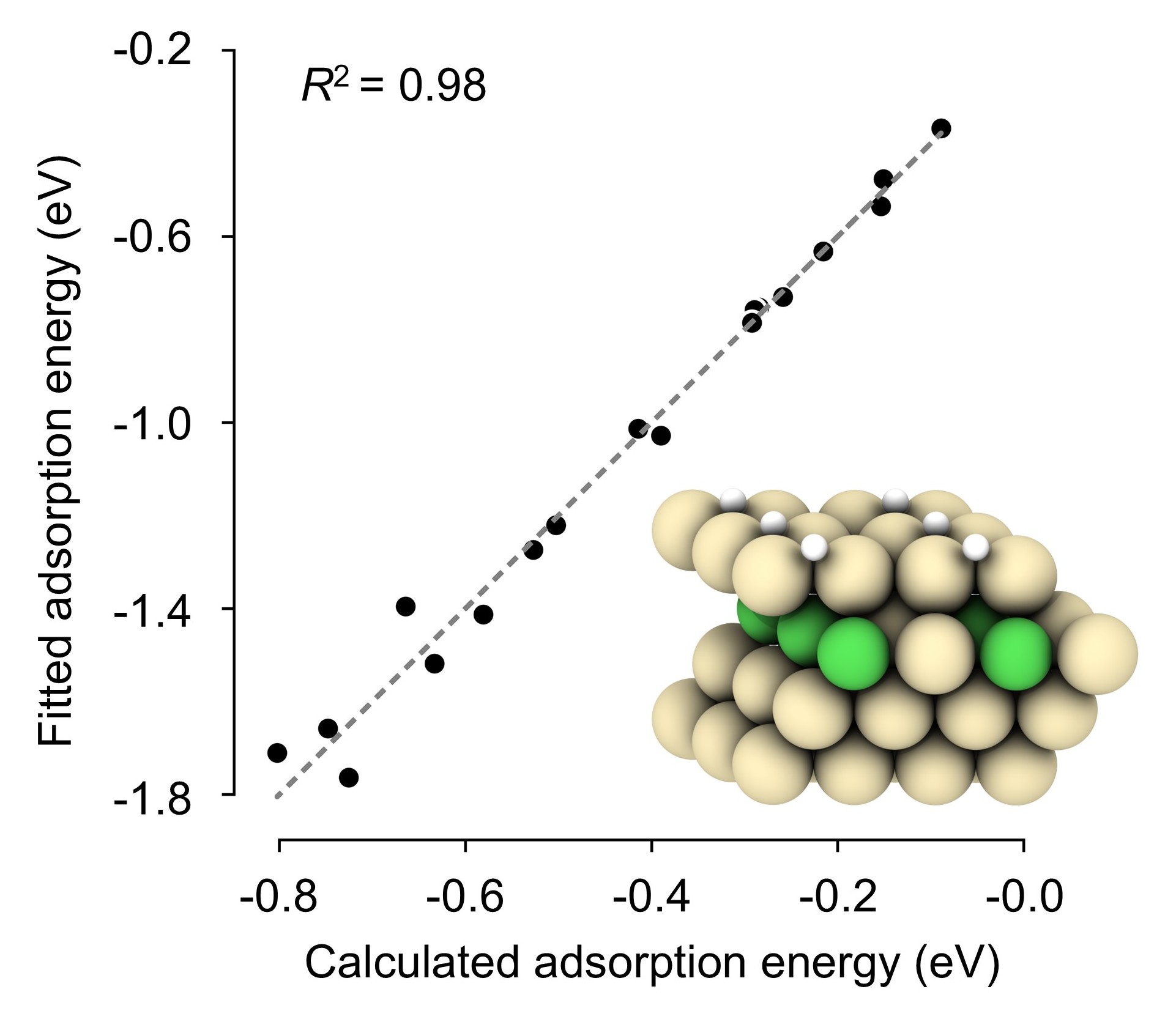}
    \caption{Parity plot comparing model-predicted adsorption energies with first-principles calculations, along with a representative atomic configuration with 6/12 coverage for $\mathrm{Pt_3Ni(111)}$.}
    \label{fig5}
\end{figure}

The model was then extended to model hydrogen adsorption on the PtNi alloy surface, where two distinct adsorption site types are considered ({Fig.~S5}). These sites differ in their local chemical environment—one having 2 Ni atoms and 1 Pt atom in the second layer (energetically favorable), and the other having 1 Ni atom and 2 Pt atoms in the second layer (less favorable). This introduces heterogeneity in adsorption behavior, which is captured through separate coverages, denoted as $\theta_1$ and $\theta_2$, for the two site types. $\theta_1 = {N_{1}}/{N_\text{s}}$ and $\theta_2 = {N_{2}}/{N_\text{s}}$ represent the fractional coverage of hydrogen on sites of type 1 and type 2, respectively. The energy of the system, $E$, is defined as
\begin{multline} 
E = I_{1} N_{1} + I_{2} N_{2} +\frac{1}{2} J_{11} N_{1} z_{11} \theta_1 + \frac{1}{2} J_{22} N_{2} z_{22} \theta_2 \\+\frac{1}{2}J_{12} N_{1} z_{12} \theta_2 +\frac{1}{2} J_{21} N_{2} z_{21} \theta_1.
\label{eq:E_ptni} 
\end{multline}
Equation \ref{eq:E_ptni} includes not only the site-specific adsorption energies $I_{1}$ and $I_{2}$, but also three distinct pairwise hydrogen interaction terms: $J_{11}$ (interactions between H atoms at site 1), $J_{22}$ (at site 2), and $J_{12}$ (interactions between H atoms occupying adjacent sites of different types). These terms account for lateral repulsion or attraction and are weighted by the number of nearest neighbors and coverage. Eighteen different adsorption configurations (Fig.~S6) were evaluated, encompassing various combinations of low- and high-energy adsorption sites and a range of coverages (Fig.~\ref{fig5}). The fcc site is used since it is the most stable adsorption site ({Fig.~S2b}). These configurations were used to fit Eq.~\ref{eq:E_ptni}, which incorporates three distinct lateral interaction parameters to capture the mixed-site environment. The resulting fitted energies are compared with first-principles data in Fig.~\ref{fig5}, demonstrating that the model accurately captures the more complex adsorption landscape of alloyed surfaces ($R^2=0.98$).

\subsubsection{Adsorption Isotherm for Alloys}
For the case of bimetallic surfaces, configurational entropy, $\Delta S_{\text{conf}}(\theta)$, is generalized to include independent contributions from each site type. This mean-field expression reflects the combinatorial entropy associated with distributing hydrogen atoms over a binary lattice 
\begin{multline}
\Delta S_\text{conf}(\theta) = k_{\rm B} [ \theta_1 \ln \theta_1 + (1 - \theta_1) \ln(1 - \theta_1) + \theta_2 \ln \theta_2 \\+ (1 - \theta_2) \ln(1 - \theta_2) ]
\label{eq:Sconf_ptni} 
\end{multline}
The chemical work term accounts for the exchange of protons and electrons with the electrochemical reservoir 
\begin{equation} \sum \mu N = N_{1} \,\mu_{{\rm H}^+} + N_1 \,\mu_{e^-} + N_{2} \,\mu_{{\rm H}^+} + N_{2} \,\mu_{e^-} \label{eq:mu_sum_ptni} \end{equation}
Minimizing the grand potential with respect to the number of adsorbed hydrogen atoms on each site, at $\mathrm{pH}=0$, yields the equilibrium condition: 
\begin{multline} \left( \frac{\partial \Omega}{\partial N_{1}} \right) = I_{1} + \frac{z_{11}}{2} J_{11} \left( \theta_1 + \frac{N_{1}}{N_\text{s}} \right) + \frac{z_{12}}{2} J_{12} \theta_2 \\
+\frac{z_{21}}{2} J_{21} \theta_2 + k_{\rm B} T \ln \left( \frac{\theta_1}{1 - \theta_1} \right) + e(U-U_{\rm SHE}) = 0 \label{eq:dOmega_dNH1} \end{multline}
By grouping terms, the resulting set of coupled equations for equilibrium coverage on each site can be written as 
\begin{eqnarray}
I_{1} + z_{11} J_{11} \theta_1 + z_{12} J_{12} \theta_2 + k_{\rm B} T \ln \left( \frac{\theta_1}{1 - \theta_1} \right) \nonumber \\ = - e(U-U_{\rm SHE}) \\ I_{2} + z_{22} J_{22} \theta_2 + z_{12} J_{12} \theta_1 + k_{\rm B} T \ln \left( \frac{\theta_2}{1 - \theta_2} \right) \nonumber \\  = - e(U-U_{\rm SHE})
\label{eq:final_PtM}
\end{eqnarray}
These equations link the applied potential $U$ to the site-specific coverages $\theta_1$ and $\theta_2$ 
which are bounded by 
\begin{equation}
0 \leq \theta_1 \leq 1, \qquad 0 \leq \theta_2 \leq 1. \label{eq:theta_bounds} 
\end{equation}
Solving this coupled system as a function of $U$ yields the potential-dependent hydrogen coverage on each site. To assess the overall hydrogen occupation on the PtNi surface, the total coverage $\theta_{\rm total}$ is defined as the fraction of all adsorption sites occupied by hydrogen, combining contributions from both type-1 (e.g., Pt-rich) and type-2 (e.g., Ni-rich) sites. Given the site-specific coverages 
$\theta_{1}$ and $\theta_{2}$, and assuming both site types are equally represented (i.e., a 1:1 ratio), the net coverage is calculated as the average of the two:
\begin{equation} 
\theta =\frac{\theta_1 + \theta_2}{2}. 
\label{eq:theta_total} 
\end{equation}

The performance of our model for bimetallic surfaces is evaluated against the experimental results of R. Stamenkovic \textit{et al.}~\citep{stamenkovic2007improved} for the $\mathrm{Pt_3Ni(111)}$ surface. The calculated coverage is shown in Fig.~\ref{fig3} which shows good agreement with experimental data. Comparing between the experimental hydrogen coverages for Pt(111) and $\mathrm{Pt_3Ni(111)}$, shown in Fig.~\ref{fig3} reveals a significantly lower total coverage for the alloy surface. This trend is well captured by our model and can be attributed to the thermodynamic unfavorability of hydrogen occupying the higher-energy adsorption sites within the studied potential window. These sites remain largely unoccupied, resulting in a lower overall hydrogen coverage compared to Pt(111).  These results demonstrate the model’s capability to capture the electrochemical behavior of alloy surfaces and highlight its potential utility for screening new HER catalyst candidates. The computed coverage and CV curves for all the alloys studied are shown in {Figs.~S7 to S9} in the Supplementary Information section.

\begin{figure*}[ht]
    \centering
    \includegraphics[width=0.8\textwidth]{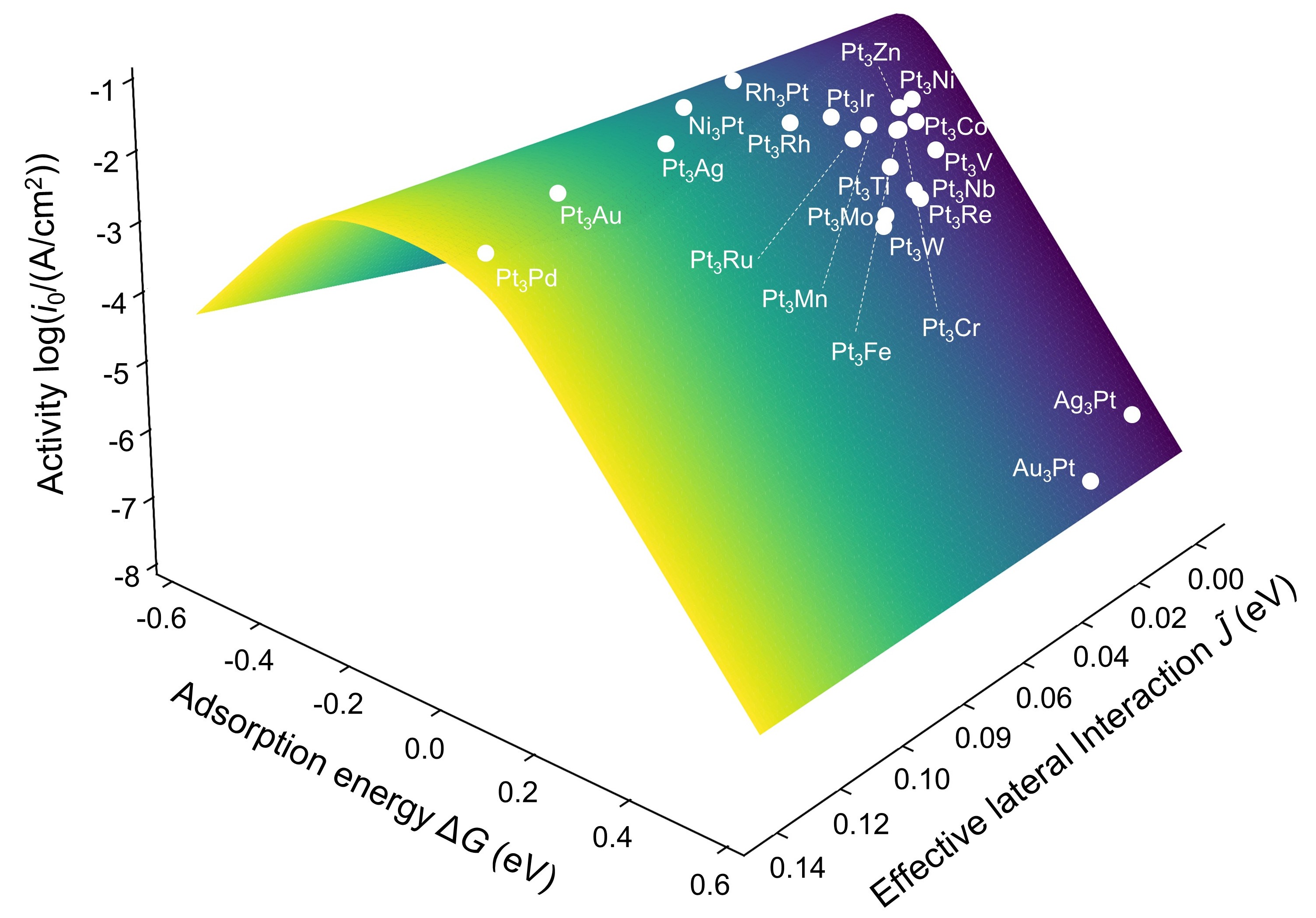}
    \caption{Volcano ridge representation for HER catalysts, illustrating catalytic activity as a function of adsorption energy and effective lateral interactions. This three-dimensional framework captures the coupled influence of adsorption energy and interaction strength on catalytic performance. The color scale corresponds to the effective lateral interaction parameter, $\tilde{J}_{11}$. All activities are calculated using a transfer coefficient of $\alpha = 0.5$.}
    \label{fig6}
\end{figure*}

\begin{figure*}[ht]
    \centering
    \includegraphics[width=\textwidth]{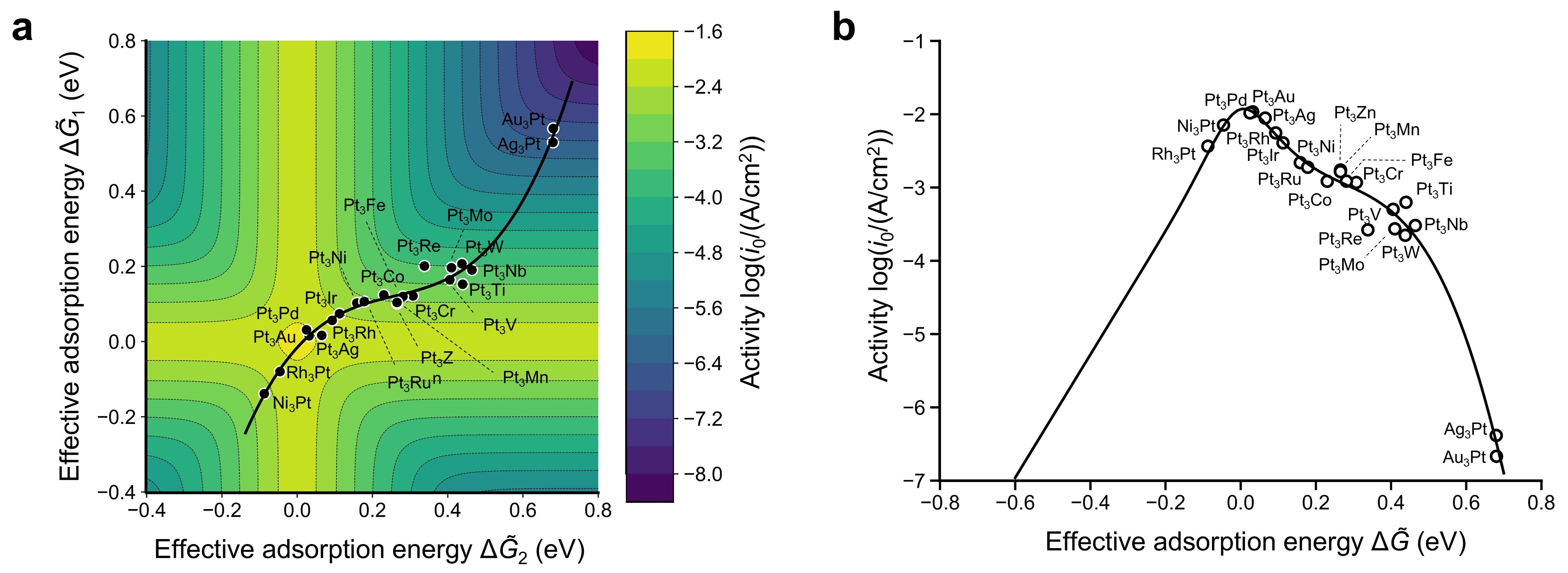}
    \caption{\textbf{a} Activity map of the investigated alloys represented as a function of the effective adsorption free energies of two distinct sites, $\Delta \tilde{G}_{1}$ and $\Delta \tilde{G}_{2}$. This map highlights the non-linear relationship between site-specific descriptors. \textbf{b} Corresponding multi-peaked volcano plot obtained by mapping the multidimensional activity landscape onto a single effective descriptor ($\Delta \tilde{G}$). While this projection can, in principle, be performed along either $\Delta \tilde{G}_{1}$ or $\Delta \tilde{G}_{2}$, the choice of $\Delta \tilde{G}_{2}$ is arbitrary and does not affect the qualitative features of the resulting activity trends. This representation captures the combined influence of adsorption energetics and lateral interactions across different sites, revealing multiple activity peaks.}
    \label{fig7}
\end{figure*}  

\subsubsection{Volcano Plots for Alloys}

For alloys, the challenge of finding the right descriptor is even harder since we have multiple adsorption sites with different energetics.  To simplify the volcano plot for alloys, Fig.~\ref{fig6} considers the most favorable adsorption site and the interactions within that site. Analysis of the site occupation behavior ({Fig.~S10}) shows that when the adsorption energies are comparable, both sites become populated, and cross-interaction terms contribute to the overall behavior. Conversely, when the difference in adsorption energies between sites is large, the less favorable site remains largely unoccupied, and its contribution to activity becomes negligible. Nevertheless, $J_{11}$ remains the dominant interaction term across all regimes because it describes the lateral interactions on the most favorable site, which is consistently occupied. These observations highlight the dominant role of $J_{11}$ relative to $J_{12}$ and $J_{22}$, motivating our decision to simplify the model by considering only the adsorption and lateral interactions of the most favorable site. {Figure S11} further demonstrates that including all interaction parameters versus including only those associated with the most favorable site results in only minor deviations in predicted activity trends. 

Rather than neglecting the contribution of secondary sites entirely, their effect is incorporated implicitly through an effective interaction parameter, enabling a reduced yet physically consistent description. We have constructed a 3D representation of the volcano (volcano ridge) where the y-axis is the adsorption energy ($\Delta G_{\rm H^*}$) and the x-axis is the effective lateral interaction ($\tilde{J}_{11}$), which is derived from Eq.~\ref{eq:final_PtM} by grouping the interaction contributions associated with the dominant site:
\begin{equation}
\begin{aligned}
I_{1} + z_{11} \theta_{1} J_{11} + z_{12} \theta_{2} J_{12}
&= I_{1} + z_{11} \theta_{1} \tilde{J}_{11} \\
\rightarrow \tilde{J}_{11}
&\equiv z_{11} \left( J_{11} + \frac{z_{12}}{z_{11}} J_{12} \frac{\theta_{2}}{\theta_{1}} \right).
\end{aligned}
\label{J11_eff}
\end{equation}
This expression follows from a projection of the interaction contributions onto the dominant adsorption site, where cross-site interactions are incorporated through their influence on relative site occupancies. The resulting effective interaction parameter captures not only the intrinsic interaction strength between adsorbates occupying the most favorable sites ($J_{11}$), but also the indirect influence of neighboring, less favorable sites through cross-interaction terms ($J_{12}$), weighted by their relative occupancies. This effective interaction parameter therefore provides a physically meaningful measure of the net interaction environment experienced by adsorbates on the dominant site.

Within this three-dimensional volcano, catalytic activity is no longer confined to a single curve but instead spans a continuous surface, forming a volcano ridge. This representation reveals that optimal activity can be achieved through multiple combinations of adsorption energy and interaction strength, rather than a single optimal binding energy as implied by conventional volcano models. In particular, systems with suboptimal adsorption energies can still exhibit high activity if compensated by favorable interaction effects, which cooperatively modify surface coverage and reaction kinetics. This expanded design space provides a natural explanation for experimentally observed improvements in catalytic performance upon alloying, even when shifts in adsorption energy alone would not predict such enhancements. We found that alloying Pt with Au, Pd, and Ag would result in better activity for HER which has been observed experimentally \citep{fu2023continuous, rakovcevic2021ptau, sun2024revisiting, wu2007high, bhalothia2020high, jebaslinhepzybai2020facile, liu2018one, huang2020ptag, weng2017simple}. 

Although the model can be simplified by considering a single adsorption site, our objective is to develop a descriptor that does not rely on such simplifications and remains applicable to systems of arbitrary complexity. To this end, we construct an activity map in Fig.~\ref{fig7}a, where the axes correspond to the effective adsorption free energies of two distinct sites, $\Delta \tilde{G}_{1}$ and $\Delta \tilde{G}_{2}$:
\begin{equation}
\Delta \tilde{G}_{1}
= I_{1} + z_{11} \theta_{1} J_{11} + z_{12} \theta_{2} J_{12} + 0.24 \ \text{eV},
\label{dG1}
\end{equation}
\begin{equation}
\Delta \tilde{G}_{2}
= I_{2} + z_{22} \theta_{2} J_{22} + z_{12} \theta_{1} J_{12} + 0.24 \ \text{eV}.
\label{dG2}
\end{equation}
These descriptors incorporate both intrinsic adsorption energetics and interaction-induced corrections arising from site-specific coverage, thereby reflecting the full thermodynamic state of the surface under operating conditions. Importantly, the coverages $\theta_1$ and $\theta_2$ are not independent variables, but are instead coupled through the equilibrium conditions governing adsorption. As shown in Fig.~\ref{fig7}a, this coupling gives rise to a nonlinear correlation between the two descriptors. Consequently, the catalytic activity of multi-site systems cannot, in general, be represented by a simple single-peaked volcano relationship. Instead, the activity landscape becomes inherently multidimensional, with the possibility of multiple local maxima corresponding to different regimes of site occupancy and interaction balance.

To enable a reduced representation of this multidimensional landscape, we introduce a mapping between the two descriptors by fitting a third-order polynomial function between them for each system. The choice of a third-order form represents the lowest-order polynomial capable of capturing the observed curvature and inflection behavior in the descriptor–descriptor relationship, while avoiding overfitting. Importantly, this mapping is not intended as a purely empirical fit, but rather as a systematic projection of the constrained thermodynamic manifold defined by the coupled coverage equations onto a one-dimensional coordinate. Using this mapping, the total catalytic activity can be expressed as a function of a single descriptor, $\Delta \tilde{G}_{2}$, as
\begin{equation}
i_{1+2}(\Delta \tilde{G}_{2})=i_1(\Delta\tilde{G}_{1}(\Delta \tilde{G}_{2}))+i_2(\Delta \tilde{G}_{2}),
\label{activity_total}
\end{equation}
where $\Delta\tilde{G}_{1}(\Delta \tilde{G}_{2})$ stands fro the fitted polynomial relationship shown in Fig.~\ref{fig7}a, $i_1$ and $i_2$ are the site-specific activities, and $i_{1+2}$ represents the total activity of the system. The multi-peaked volcano shown in Fig.~\ref{fig7}b is constructed as a function of $\Delta \tilde{G}_{2}$ via Eq.~\ref{activity_total}.  This reduced representation demonstrates that it is possible to project the inherently high-dimensional activity landscape of multi-site catalytic systems onto a single effective descriptor without discarding the essential physics governing adsorption and interaction effects. As a result, it retains the conceptual simplicity of Sabatier’s principle while extending its applicability to complex, multi-site catalytic alloys.

\section{Discussion}

In this work, we developed an analytical framework that explicitly accounts for lateral interactions and applied it to model hydrogen coverage on Pt-based alloy surfaces. By combining density-functional theory calculations with statistical thermodynamics, we establish a quantitative connection between adsorption energetics, coverage effects, and experimentally measurable electrochemical responses. The model is rigorously validated through direct comparison with experimental coverage isotherms for Pt(111) and $\mathrm{Pt_3Ni(111)}$ surfaces, demonstrating its predictive capability. We show that lateral interactions play a decisive role in governing hydrogen coverage. Specifically, attractive interactions lead to steeper coverage-potential curves, while repulsive interactions produce a more gradual evolution of surface coverage with applied potential.

Furthermore, we demonstrate that lateral interactions fundamentally reshape volcano-type activity trends and must be explicitly considered in electrocatalytic modeling. Their inclusion leads to systematic distortions of the conventional volcano profile, including both broadening and shifts in the activity maximum relative to models that neglect adsorbate-adsorbate effects. 
Depending on their magnitude and sign, these interactions can either flatten or sharpen the activity peak, thereby altering both the activity landscape and the predictions of optimal catalyst compositions.

Building on these insights, we move beyond conventional two-dimensional volcano plots and introduce a three-dimensional "volcano ridge" representation, in which catalytic activity is described as a function of both the adsorption free energy and the effective lateral interaction strength. This unified framework captures the coupled effects of adsorption energetics, coverage, and interactions, providing a more complete and physically grounded descriptor space for alloy electrocatalysts.

Finally, we address the challenge of describing multi-site catalytic systems without resorting to simplifying assumptions based on a single adsorption site. By constructing an activity map in terms of the effective adsorption energies of multiple sites, we show that the relationship between site-specific descriptors is inherently non-linear, precluding a simple single-peaked volcano description. To overcome this limitation, we introduce a mapping between site descriptors using a polynomial representation, enabling the construction of a multi-peaked volcano relationship that captures the combined contributions of multiple adsorption sites. This approach effectively reduces the high-dimensional activity landscape to a single descriptor while retaining the essential physics of site heterogeneity and lateral interactions. As a result, it preserves the conceptual simplicity of Sabatier’s principle while extending its applicability to catalytic alloys of arbitrary complexity, providing a powerful framework for the design of next-generation electrocatalysts.

\section{Methods}
\subsection{Hydrogen Evolution Reaction}
The electrochemical process of hydrogen adsorption on a metal surface is characterized as a reaction between protons from the electrolyte and electrons from an electrode maintained at a constant potential $U$, which is described by the following reaction
\begin{equation}
* + {\rm H}^+_{\rm (aq)} + e^- \leftrightarrows {\rm H}^*,
\label{reaction1}
\end{equation}
where $*$ indicates an adsorption site and H$^*$ denotes a hydrogen atom adsorbed on the metal surface. The protons are assumed to be in equilibrium with the aqueous phase, and their chemical potential, $\mu_{{\rm H}^+}$, is controlled by the pH of the electrolyte, following
\begin{equation}
\mu_{{\rm H}^+} = \mu_{{\rm H}^+}^\circ - k_{\rm B}T \ln(10)\,\mathrm{pH}. 
\label{muH+}
\end{equation}
The chemical potential of the protons is referenced to the standard hydrogen electrode, where protons and molecular hydrogen reach equilibrium under standard conditions and $\mathrm{pH} = 0$, at $U = U_{\rm SHE} = 4.44\,\mathrm{V}$. The equilibrium equation reads
\begin{equation}
{\rm H}^+_{\rm (aq)} + e^- \leftrightarrows \frac{1}{2} {\rm H}_2^{\rm (g)}, 
\label{reaction2}
\end{equation}
leading to the following expression of the chemical potential for protons:
\begin{equation}
\mu_{{\rm H}^+} = eU_{\rm SHE} + \frac{1}{2} \mu_{\rm H_2}^\circ - k_{\rm B}T \ln(10)\,\mathrm{pH}, 
\label{muH+SHE}
\end{equation}
where $\mu_{\rm H_2}^\circ$, the chemical potential of molecular hydrogen in its standard state, is estimated from DFT calculations of the total energy of molecular hydrogen in vacuum
\begin{equation} 
\mu_{\rm H_2}^\circ = E_{{\rm H}_2}.
\label{eq:E_H2_dft} 
\end{equation}

\subsection{First-Principles Calculations}
Electronic-structure calculations were performed using the {\sc quantum espresso} suite for plane-wave materials simulations~\citep{giannozzi2009quantum, giannozzi2017advanced}. A four-layer, periodic 3×4 supercell with 10 angstroms of vacuum was employed to model the Pt(111) and PtM(111) surfaces. Projector-augmented-wave pseudopotentials from the {\sc pseudodojo} library were used to represent interactions with ionic cores \citep{jollet2014generation}, and the \linebreak Perdew–Burke–Ernzerhof (PBE)~\citep{perdew1996generalized} exchange-correlation functional was applied to calculate the energies. The plane wave expansions of wavefunctions and electronic charge density were truncated at kinetic energy cutoffs of 80 Ry and 320 Ry, respectively. For Brillouin zone sampling, the ${\boldsymbol k}$-point density was set to 0.025 \AA$^{-1}$. Electronic occupations were smoothed using the Marzari–Vanderbilt cold smearing method~\citep{marzari1999thermal}, with a smearing width of 0.01 Ry. All the calculations were spin-polirized to account for magnetization. Atomic positions were relaxed until total energy and forces were converged to within 1 meV per atom and 1 meV/\AA, respectively.

\section{Acknowledgments}

This work was primarily supported by the U.S. Department of Energy, Office of Science, Basic Energy Sciences, CPIMS (Condensed Phase and Interfacial Molecular Science) Program, under Award No.~DE-SC0018646. S.G.~acknowledges support from the U.S. Department of Energy, Office of Science, Office of Basic Energy Sciences under Award no. DE-SC0023415 (Center for Electrochemical Dynamics for Reactions on Surfaces, an Energy Frontier Research Center.) A subset of the first-principles calculations was carried out on Anvil at Purdue University through allocation CHM230047 from the Advanced Cyberinfrastructure Coordination Ecosystem: Services \& Support (ACCESS) program, which is supported by National Science Foundation grants nos. 2138259, 2138286, 2138307, 2137603, and 2138296. The authors gratefully acknowledge Prof. S. B. Sinnott for funding acquisition and support of this research.

\bibliography{bibliography.bib}

@article{stamenkovic2007trends,
  title={Trends in electrocatalysis on extended and nanoscale Pt-bimetallic alloy surfaces},
  author={Stamenkovic, Vojislav R and Mun, Bongjin Simon and Arenz, Matthias and Mayrhofer, Karl JJ and Lucas, Christopher A and Wang, Guofeng and Ross, Philip N and Markovic, Nenad M},
  journal={Nature materials},
  volume={6},
  number={3},
  pages={241--247},
  year={2007},
  publisher={Nature Publishing Group UK London}
}

@book{Sabatier1913,
  author    = {Sabatier, Paul},
  title     = {La Catalyse en Chimie Organique},
  year      = {1913},
  publisher = {Librairie Polytechnique},
  address   = {Paris}
}

@article{Sabatier1911HydrognationsED,
  title={Hydrog{\'e}nations et d{\'e}shydrog{\'e}nations par catalyse},
  author={Paul Sabatier},
  journal={European Journal of Inorganic Chemistry},
  year={1911},
  volume={44},
  pages={1984-2001},
  url={https://api.semanticscholar.org/CorpusID:95404799}
}

@article{giannozzi2009quantum,
  title={{QUANTUM ESPRESSO: a modular and open-source software project for quantum simulations of materials}},
  author={Giannozzi, Paolo and Baroni, Stefano and Bonini, Nicola and Calandra, Matteo and Car, Roberto and Cavazzoni, Carlo and Ceresoli, Davide and Chiarotti, Guido L and Cococcioni, Matteo and Dabo, Ismaila and others},
  journal={Journal of Physics: Condensed Matter},
  volume={21},
  number={39},
  pages={395502},
  year={2009},
  publisher={IOP Publishing}
}

@article{giannozzi2017advanced,
  title={{Advanced capabilities for materials modelling with Quantum ESPRESSO}},
  author={Giannozzi, Paolo and Andreussi, Oliviero and Brumme, Thomas and Bunau, Oana and Nardelli, M Buongiorno and Calandra, Matteo and Car, Roberto and Cavazzoni, Carlo and Ceresoli, Davide and Cococcioni, Matteo and others},
  journal={Journal of Physics: Condensed Matter},
  volume={29},
  number={46},
  pages={465901},
  year={2017},
  publisher={IOP Publishing}
}

@article{jollet2014generation,
  title={{Generation of Projector Augmented-Wave atomic data: A 71 element validated table in the XML format}},
  author={Jollet, Fran{\c{c}}ois and Torrent, Marc and Holzwarth, Natalie},
  journal={Computer Physics Communications},
  volume={185},
  number={4},
  pages={1246--1254},
  year={2014},
  publisher={Elsevier}
}

@article{perdew1996generalized,
  title={{Generalized gradient approximation made simple}},
  author={Perdew, John P and Burke, Kieron and Ernzerhof, Matthias},
  journal={Physical Review Letters},
  volume={77},
  number={18},
  pages={3865},
  year={1996},
  publisher={APS}
}

@article{marzari1999thermal,
  title={{Thermal contraction and disordering of the Al (110) surface}},
  author={Marzari, Nicola and Vanderbilt, David and De Vita, Alessandro and Payne, MC},
  journal={Physical Review Letters},
  volume={82},
  number={16},
  pages={3296},
  year={1999},
  publisher={APS}
}

@article{norskov2005trends,
  title={Trends in the exchange current for hydrogen evolution},
  author={N{\o}rskov, Jens Kehlet and Bligaard, Thomas and Logadottir, Ashildur and Kitchin, JR and Chen, Jingguang G and Pandelov, S and Stimming, UJJoTES},
  journal={Journal of The Electrochemical Society},
  volume={152},
  number={3},
  pages={J23},
  year={2005},
  publisher={IOP Publishing}
}

@article{stamenkovic2007improved,
  title={Improved oxygen reduction activity on Pt3Ni (111) via increased surface site availability},
  author={Stamenkovic, Vojislav R and Fowler, Ben and Mun, Bongjin Simon and Wang, Guofeng and Ross, Philip N and Lucas, Christopher A and Markovic, Nenad M},
  journal={science},
  volume={315},
  number={5811},
  pages={493--497},
  year={2007},
  publisher={American Association for the Advancement of Science}
}

@article{fu2023continuous,
  title={Continuous-flow electrosynthesis of ammonia by nitrogen reduction and hydrogen oxidation},
  author={Fu, Xianbiao and Pedersen, Jakob B and Zhou, Yuanyuan and Saccoccio, Mattia and Li, Shaofeng and Sa{\v{z}}inas, Rokas and Li, Katja and Andersen, Suzanne Z and Xu, Aoni and Deissler, Niklas H and others},
  journal={Science},
  volume={379},
  number={6633},
  pages={707--712},
  year={2023},
  publisher={American Association for the Advancement of Science}
}

@article{qiao2021pt3fe,
  title={Pt3Fe nanoparticles on B, N-codoped carbon as oxygen reduction and pH-universal hydrogen evolution electrocatalysts},
  author={Qiao, Yueyang and Cui, Jiayao and Qian, Fangren and Xue, Xiaoyi and Zhang, Xiangping and Zhang, Haitao and Liu, Wenqi and Li, Xiaojin and Chen, Qingjun},
  journal={ACS Applied Nano Materials},
  volume={5},
  number={1},
  pages={318--325},
  year={2021},
  publisher={ACS Publications}
}

@article{zhong2017double,
  title={Double nanoporous structure with nanoporous PtFe embedded in graphene nanopores: highly efficient bifunctional electrocatalysts for hydrogen evolution and oxygen reduction},
  author={Zhong, Xing and Wang, Lei and Zhuang, Zhenzhan and Chen, Xianlang and Zheng, Jian and Zhou, Yulin and Zhuang, Guilin and Li, Xiaonian and Wang, Jianguo},
  journal={Advanced Materials Interfaces},
  volume={4},
  number={5},
  pages={1601029},
  year={2017},
  publisher={Wiley Online Library}
}

@article{lin2020structurally,
  title={Structurally ordered Pt3Co nanoparticles anchored on N-doped graphene for highly efficient hydrogen evolution reaction},
  author={Lin, Caoxin and Huang, Zhiqiao and Zhang, Zeyi and Zeng, Tang and Chen, Runzhe and Tan, Yangyang and Wu, Wei and Mu, Shichun and Cheng, Niancai},
  journal={ACS Sustainable Chemistry \& Engineering},
  volume={8},
  number={45},
  pages={16938--16945},
  year={2020},
  publisher={ACS Publications}
}

@article{zhang2021engineering,
  title={Engineering platinum--cobalt nano-alloys in porous nitrogen-doped carbon nanotubes for highly efficient electrocatalytic hydrogen evolution},
  author={Zhang, Song Lin and Lu, Xue Feng and Wu, Zhi-Peng and Luan, Deyan and Lou, Xiong Wen},
  journal={Angewandte Chemie International Edition},
  volume={60},
  number={35},
  pages={19068--19073},
  year={2021},
  publisher={Wiley Online Library}
}

@article{yu2023high,
  title={High-density frustrated lewis pair for high-performance hydrogen evolution},
  author={Yu, Wenhao and Zhang, Yanyun and Qin, Yingnan and Zhang, Dan and Liu, Kang and Bagliuk, GA and Lai, Jianping and Wang, Lei},
  journal={Advanced Energy Materials},
  volume={13},
  number={2},
  pages={2203136},
  year={2023},
  publisher={Wiley Online Library}
}

@article{zhang2019bimetallic,
  title={Bimetallic PtCo alloyed nanodendritic assemblies as an advanced efficient and robust electrocatalyst for highly efficient hydrogen evolution and oxygen reduction},
  author={Zhang, Xiao-Fang and Meng, Han-Bin and Chen, Hong-Yan and Feng, Jiu-Ju and Fang, Ke-Ming and Wang, Ai-Jun},
  journal={Journal of Alloys and Compounds},
  volume={786},
  pages={232--239},
  year={2019},
  publisher={Elsevier}
}

@article{zhang2021wox,
  title={WOx-surface decorated PtNi@ Pt dendritic nanowires as efficient pH-universal hydrogen evolution electrocatalysts},
  author={Zhang, Weiyu and Huang, Bolong and Wang, Kai and Yang, Wenxiu and Lv, Fan and Li, Na and Chao, Yuguang and Zhou, Peng and Yang, Yong and Li, Yingjie and others},
  journal={Advanced Energy Materials},
  volume={11},
  number={3},
  pages={2003192},
  year={2021},
  publisher={Wiley Online Library}
}

@article{mondal2022morphology,
  title={Morphology-Tuned Pt3Ge Accelerates Water Dissociation to Industrial-Standard Hydrogen Production over a wide pH Range},
  author={Mondal, Soumi and Sarkar, Shreya and Bagchi, Debabrata and Das, Tisita and Das, Risov and Singh, Ashutosh Kumar and Prasanna, Ponnappa Kechanda and Vinod, CP and Chakraborty, Sudip and Peter, Sebastian C},
  journal={Advanced Materials},
  volume={34},
  number={30},
  pages={2202294},
  year={2022},
  publisher={Wiley Online Library}
}

@article{pang2022laser,
  title={Laser-assisted high-performance PtRu alloy for pH-universal hydrogen evolution},
  author={Pang, Beibei and Liu, Xiaokang and Liu, Tianyang and Chen, Tao and Shen, Xinyi and Zhang, Wei and Wang, Sicong and Liu, Tong and Liu, Dong and Ding, Tao and others},
  journal={Energy \& Environmental Science},
  volume={15},
  number={1},
  pages={102--108},
  year={2022},
  publisher={Royal Society of Chemistry}
}

@article{chen2022ptco,
  title={PtCo@ PtSn heterojunction with high stability/activity for pH-universal H2 evolution},
  author={Chen, Jinli and Qian, Guangfu and Zhang, Hao and Feng, Shouquan and Mo, Yanshan and Luo, Lin and Yin, Shibin},
  journal={Advanced Functional Materials},
  volume={32},
  number={5},
  pages={2107597},
  year={2022},
  publisher={Wiley Online Library}
}

@article{ni2023construction,
  title={Construction of hierarchical and self-supported NiFe-Pt3Ir electrode for hydrogen production with industrial current density},
  author={Ni, Zhenrui and Luo, Cheng and Cheng, Bei and Kuang, Panyong and Li, Youji and Yu, Jiaguo},
  journal={Applied Catalysis B: Environmental},
  volume={321},
  pages={122072},
  year={2023},
  publisher={Elsevier}
}

@article{chen2024restructuring,
  title={Restructuring the interfacial active sites to generalize the volcano curves for platinum-cobalt synergistic catalysis},
  author={Chen, Wenyao and Shi, Yao and Liu, Changwei and Ren, Zhouhong and Huang, Zikun and Chen, Zhou and Zhang, Xiangxue and Liang, Shanshan and Xie, Lei and Lian, Cheng and others},
  journal={Nature Communications},
  volume={15},
  number={1},
  pages={8995},
  year={2024},
  publisher={Nature Publishing Group UK London}
}

@article{liang2024unravelling,
  title={Unravelling the effects of active site density and energetics on the water oxidation activity of iridium oxides},
  author={Liang, Caiwu and Rao, Reshma R and Svane, Katrine L and Hadden, Joseph HL and Moss, Benjamin and Scott, Soren B and Sachs, Michael and Murawski, James and Frandsen, Adrian Malthe and Riley, D Jason and others},
  journal={Nature Catalysis},
  volume={7},
  number={7},
  pages={763--775},
  year={2024},
  publisher={Nature Publishing Group UK London}
}

@article{ke2022three,
  title={Three-dimensional activity volcano plot under an external electric field},
  author={Ke, Changming and Lin, Zijing and Liu, Shi},
  journal={ACS Catalysis},
  volume={12},
  number={21},
  pages={13542--13548},
  year={2022},
  publisher={ACS Publications}
}

@article{grabow2010understanding,
  title={Understanding trends in catalytic activity: the effect of adsorbate--adsorbate interactions for CO oxidation over transition metals},
  author={Grabow, Lars C and Hvolb{\ae}k, Britt and N{\o}rskov, Jens K},
  journal={Topics in Catalysis},
  volume={53},
  number={5},
  pages={298--310},
  year={2010},
  publisher={Springer}
}

@article{wang2025revisiting,
  title={Revisiting Volcano Plots: An Analytical 3D Approach},
  author={Wang, Yifan and Leng, Yecheng and Tu, Wenguang and Zou, Zhigang and Zhu, Xi},
  journal={CCS Chemistry},
  pages={1--10},
  year={2025},
  publisher={Chinese Chemical Society Zhongguancun, Haidian, Beijing 100190, China}
}

@article{qi2012adsorbate,
  title={Adsorbate interactions on surface lead to a flattened Sabatier volcano plot in reduction of oxygen},
  author={Qi, Liang and Li, Ju},
  journal={Journal of Catalysis},
  volume={295},
  pages={59--69},
  year={2012},
  publisher={Elsevier}
}

@article{rakovcevic2021ptau,
  title={PtAu nanoparticles supported by reduced graphene oxide as a highly active catalyst for hydrogen evolution},
  author={Rako{\v{c}}evi{\'c}, Lazar and Simatovi{\'c}, Ivana Stojkovi{\'c} and Maksi{\'c}, Aleksandar and Raji{\'c}, Vladimir and {\v{S}}trbac, Svetlana and Sreji{\'c}, Irina},
  journal={Catalysts},
  volume={12},
  number={1},
  pages={43},
  year={2021},
  publisher={MDPI}
}

@article{sun2024revisiting,
  title={Revisiting the activity origin of the PtAu 24 (SR) 18 nanocluster for enhanced electrocatalytic hydrogen evolution by combining first-principles simulations with the experimental in situ FTIR technique},
  author={Sun, Fang and Qin, Lubing and Tang, Zhenghua and Tang, Qing},
  journal={Chemical Science},
  volume={15},
  number={39},
  pages={16142--16155},
  year={2024},
  publisher={Royal Society of Chemistry}
}

@article{wu2007high,
  title={High activity PtPd-WC/C electrocatalyst for hydrogen evolution reaction},
  author={Wu, Mei and Shen, Pei Kang and Wei, Zidong and Song, Shuqin and Nie, Ming},
  journal={Journal of Power Sources},
  volume={166},
  number={2},
  pages={310--316},
  year={2007},
  publisher={Elsevier}
}

@article{bhalothia2020high,
  title={High-performance and stable hydrogen evolution reaction achieved by Pt trimer decoration on ultralow-metal loading bimetallic PtPd nanocatalysts},
  author={Bhalothia, Dinesh and Huang, Tzu-Hsi and Chang, Chia-Wen and Lin, Ting-Han and Wu, Shun-Chi and Wang, Kuan-Wen and Chen, Tsan-Yao},
  journal={ACS Applied Energy Materials},
  volume={3},
  number={11},
  pages={11142--11152},
  year={2020},
  publisher={ACS Publications}
}

@article{jebaslinhepzybai2020facile,
  title={Facile galvanic replacement method for porous Pd@ Pt nanoparticles as an efficient HER electrocatalyst},
  author={Jebaslinhepzybai, Balasingh Thangadurai and Prabu, Natarajan and Sasidharan, Manickam},
  journal={International Journal of Hydrogen Energy},
  volume={45},
  number={19},
  pages={11127--11137},
  year={2020},
  publisher={Elsevier}
}

@article{liu2018one,
  title={One-pot aqueous fabrication of reduced graphene oxide supported porous PtAg alloy nanoflowers to greatly boost catalytic performances for oxygen reduction and hydrogen evolution},
  author={Liu, Qing and He, Ya-Mei and Weng, Xuexiang and Wang, Ai-Jun and Yuan, Pei-Xian and Fang, Ke-Ming and Feng, Jiu-Ju},
  journal={Journal of colloid and interface science},
  volume={513},
  pages={455--463},
  year={2018},
  publisher={Elsevier}
}

@article{huang2020ptag,
  title={PtAg alloy nanoparticles embedded in polyaniline as electrocatalysts for formate oxidation and hydrogen evolution},
  author={Huang, Meng and Zhang, Hanyu and Yin, Song and Zhang, Xiaoxue and Wang, Jun},
  journal={ACS Applied Nano Materials},
  volume={3},
  number={4},
  pages={3760--3766},
  year={2020},
  publisher={ACS Publications}
}

@article{weng2017simple,
  title={Simple one-pot synthesis of solid-core@ porous-shell alloyed PtAg nanocrystals for the superior catalytic activity toward hydrogen evolution and glycerol oxidation},
  author={Weng, Xuexiang and Liu, Qing and Wang, Ai-Jun and Yuan, Junhua and Feng, Jiu-Ju},
  journal={Journal of Colloid and Interface Science},
  volume={494},
  pages={15--21},
  year={2017},
  publisher={Elsevier}
}

@article{yang2022reconciling,
  title={Reconciling the volcano trend with the Butler--Volmer model for the hydrogen evolution reaction},
  author={Yang, Timothy T and Saidi, Wissam A},
  journal={The Journal of Physical Chemistry Letters},
  volume={13},
  number={23},
  pages={5310--5315},
  year={2022},
  publisher={ACS Publications}
}

@book{bard2022electrochemical,
  title={Electrochemical methods: fundamentals and applications},
  author={Bard, Allen J and Faulkner, Larry R and White, Henry S},
  year={2022},
  publisher={John Wiley \& Sons}
}

@article{trasatti1972work,
  title={Work function, electronegativity, and electrochemical behaviour of metals: III. Electrolytic hydrogen evolution in acid solutions},
  author={Trasatti, Sergio},
  journal={Journal of Electroanalytical Chemistry and Interfacial Electrochemistry},
  volume={39},
  number={1},
  pages={163--184},
  year={1972},
  publisher={Elsevier}
}

\end{document}